\documentclass{article}

\usepackage{arxiv}

\usepackage[utf8]{inputenc} % allow utf-8 input
\usepackage[T1]{fontenc}    % use 8-bit T1 fonts

\usepackage{hyperref}       % hyperlinks
\usepackage{url}            % simple URL typesetting
\usepackage{booktabs}       % professional-quality tables
\usepackage{amsfonts}       % blackboard math symbols
\usepackage{amsmath}
\usepackage{amssymb}
\usepackage{microtype}      % microtypography
\usepackage{graphicx}
\usepackage{lipsum}
\usepackage[numbers]{natbib} % NUMERIC tarzda atıf için
\usepackage{doi}

\hypersetup{
  colorlinks = true,
  linkcolor  = black,  % section, equation vb. linkler
  citecolor  = blue,   % \cite linkleri
  urlcolor   = blue    % URL linkleri
}

\title{Atangana–Baleanu Regularized Wavelet Compression for Astronomical Time-Series}

\author{ \href{https://orcid.org/0000-0001-7402-4468}{\includegraphics[scale=0.06]{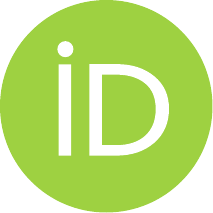}\hspace{1mm}Taylan Demir}\thanks{Use footnote for providing further
		information about author (webpage, alternative
		address)---\emph{not} for acknowledging funding agencies.} \\
	Department of Mathematics\\
	  Ankara University\\
	Ankara, Turkey \\
	\texttt{taylandemir@ankara.edu.tr} \\
	\And
	\href{https://orcid.org/0009-0005-6060-0547}{\includegraphics[scale=0.06]{orcid.pdf}\hspace{1mm}Atakan Koçyiğit} \\
	Department of Electrical and Electronics Engineering\\
	  Atılım University\\
	Ankara, Turkey \\
	\texttt{atakan.kocyigit@megetek.com.tr} \\
}

\renewcommand{\shorttitle}{\textit{arXiv} Template}

\hypersetup{
pdftitle={A template for the arxiv style},
pdfsubject={q-bio.NC, q-bio.QM},
pdfauthor={David S.~Hippocampus, Elias D.~Striatum},
pdfkeywords={First keyword, Second keyword, More},
}

\begin{document}
\maketitle
\begin{abstract}
Astronomical light curves are noisy and irregular, so compression must reduce size without erasing weak transients. We propose a fractional wavelet compression method where wavelet coefficients
are regularized via an Atangana Baleanu Caputo derivative with a nonsingular Mittag Leffler kernel. The induced long memory smoothing suppresses noise while preserving coherent transits,
flares and oscillations. We give the coefficient level formulation, an efficient implementation, and comparisons with classical discrete wavelet thresholding, showing competitive compression with
improved retention of low-amplitude events.
\end{abstract}
\section{Introduction}
Today, astronomy produces vast quantities of data over extended times and through massive data collections. The adoption of satellite missions such as NASA’s TESS has resulted in complex analyses of millions of stellar light curves and extensive public repositories of millions of temporally sequenced images taken from the TESS satellite’s long-term campaign of operations [1]. Working synergistically with TESS, other satellite-based imaging systems will produce additionally millions of light curves based upon millions of full-frame images taken from those systems [2]. Massive archival volumes and large numbers of thousands of images and records of quasi-periodic photometric variations, for example, are creating an unprecedented need for a fully functional astronomical imaging archival organization that exceeds one petabyte of data to ensure continued support of astronomical research and development efforts [3]. Each of these datasets not only tracks the locations and distances of planets, moons, stars, etc., they also show temporal changes in the observed frequency and time of occurrence of other celestial phenomena with variations on a broad range of timescales, such as planetary transits or stellar explosive and periodic transient events (e.g., eclipses). To successfully reuse the data, we must first remove a great deal of the noise in the data and recover the faint and fleeting astrophysical signals found in the light curves. This is achieved through reconstruction of the original time series from the compressed version of the data, primarily via wavelets, which provide a multi-resolution representation of the data. The DWT is particularly effective for producing a reduced-resolution representation of an astronomical light curve that allows local features in the data to be captured with only a few significant coefficients [4]. Conventional methods for compressing/de-noising astronomical images (and time series) have relied upon the concept of eliminating all coefficients with low amplitudes while preserving the larger-amplitude coefficients. This assumption of low amplitude coefficients primarily containing noise leads to a high degree of compression and the introduction of artefacts at the edges of high amplitudes; in addition, attenuation of astrophysically relevant transient (i.e., low amplitude) signals can result from this method. The issues associated with the introduction of ringing artefacts and the potential for omitting low-amplitude signals are exacerbated when long-term correlations and non-stationarity are present in the data. These additional characteristics cannot be directly represented in the typical manner. Fractional calculus presents a new way to look at memory effects using Fractional Derivatives that are made up of Non-local kernels. This allows Fractional Derivatives to represent a memory system based on how the past history has affected its present behaviour. The Atangana - Baleanu (AB) fractional derivative is one of many examples of Fractional Derivatives; AB has been studied to see what benefit it may provide. The AB is defined with a non singular Mittag-Leffler kernel and gives the ability to assign weights to historical information in such a manner that it gives a smooth transition from Exponential to Power Law Memory [5]. The manner in which to formulate the AB fractional operator eliminates the singularity problems associated with the traditional Power Law and allows a more stable Scaffold to build upon for many application fields. More recent studies have shown that the AB-type Fractional Operator can effectively model Anomalous Diffusion; Viscoelastic Behaviour; and Dynamics of Epidemic Infections: with good correlation to data collected from real-world observations; and has a clear physical Interpretation of the memory parameters [6]. Overall, Mathematical Modelling along with Numerical Simulation is the most common way to study Complex Systems, including many large scale Engineering applications. The conventional manner in which one begins to develop a model of a highly complex system is to start with the formulation of an ODE/integral-differential equations that represents the relevant physical processes. Once an ODE/integral-differential equation is established, this equation must then be discretised for the appropriate geometrical shape and time scales (using various numerical techniques). Once a model is built, the next logical step is to validate the results of the model, by comparing the results of the model to the results obtained from real world measurements or other similar sources. Because of the large number of very good numerical libraries, signal processing toolbox and visualisation libraries available in both MATLAB and Python, both environments provide for a very similar way of building/reproducing these models [4]. Therefore, in this effort, we are going to develop a model of a compression algorithm which will be based on modelling/simulating astrophysical time series. Our most important contribution to this effort is the use of fractional differential operators to provide an ongoing/regularised time evolution for the wavelet coefficient at the Cascades Level (i.e., the modelling/simulation process). Rather than applying an explicit mathematical threshold limit on the coefficients, we will instead allow the coefficients to "evolve" via solving an ordered fractional equation that incorporates both the order and the kernel parameters of the fractional equation to determine the strength of memory, as well as the time scale at which that memory exists. The final product will allow the modelling of the astrophysical time series using a process that utilises both time and scale coherence to reduce noise while preserving the time and scale coherence of the signal. With the experimental test signals and the actual light curves from space-based surveys the proposed method can achieve similar compression ratios while allowing for decreased a large quantity of artifacts, resulting in additional low-level variations. In the remainder of this paper we will provide an overview of the method by breaking down the structure into sections. Section 2 provides an introduction to the classical discrete wavelet transform method that can routinely be applied to compress astronomical data, while section 3 provides an overview of the main properties of the fractional A–B wavelet coefficient
model used to describe the way in which wavelet coefficients change over time. Section 4 describes how to use MATLAB or Python to implement the fractional A–B model numerically, and describes the datasets used to implement this model. In section 5 we compare the fractional A–B wavelet method and the classical discrete wavelet transform method to synthesized and observed time series data. Finally, section 6 summarises the paper and makes suggestions for future research that involves adaptive selection of fractional orders and possible expansion into multi-channel datasets.
\section{Background on wavelet compression}
\label{sec:wavelet_background}

\subsection{Discrete wavelet transform and filter banks}
The Discrete Wavelet Transform (DWT) provides an orthogonal or biorthogonal expansion of signals based on dilated and translated versions of the mother wavelet. Multiresolution analysis unites both the coarse and fine scale components of the expansion into nested approximation and detail spaces. Through multiresolution analysis, an entire family of signal approximations can be obtained based on progressively increasing frequency ranges, with the lowest frequency being the coarsest approximation and the highest frequency being the finest detail - [7][8]. In practice, the DWT does not require an explicit calculation of inner products involving continuous wavelets; however, the algorithm is implemented by using critically sampled filter banks. A pair of conjugate low-pass/high-pass filters performs a two-channel subband decomposition followed by two-to-one downsampling. A pyramid of subbands, representing additional levels of signal detail, can be formed by recurring iterations of the analysis stage for the low frequency branch of the filter bank. Conversely, the synthesis bank reverts the aforementioned processes (upsampling, filtering, and summing), achieving complete reconstruction when the filters of the analysis/synthesis banks fulfil the standard conditions of quadrature-mirror or biorthogonality - [9]. This algorithm also has a linear complexity (i.e., it requires a number of calculations directly related to the number of samples), allowing it to process very long one-dimensional time series sequences (i.e., produced by the modern world).

\subsection{Coefficient modelling and thresholding strategies}
WLCR has led to the development of many wavelet transforms to accommodate many signal architectures including smooth and other classes. The majority of the signal typically resides in a small number of the high-frequency Wavlet coefficients, resulting in a Comprehesive Denoising approach that models the underlying signal as the superposition of a sparse signal component and an additive noise component. Donoho and Johnstone developed a classical "soft" or "hard" thresholding method from this
approach, treating the DWT as a near diagonalisation of a class of Covariance matrices and applying shrinkage or thresholding rule to the noisy coefficients as presented in [10]. The simplest examples are the "hard" and "soft" thresholding rules, where coefficients with magnitudes less than a threshold are set to zero and coefficients above the threshold may be reduced in amplitude based on some nonlinear function. Under suitable assumptions, these methods produce nearly optimal minimax risk and adaptive rate of approximation between classes of functions in a wide range of engineering applications. Similar to the
use of hr/soft thresholding techniques, transform coders employ this same principle; when the bitstream is constructed, small coefficients are discarded or coarsely quantised and large coefficients are preserved with a higher level of accuracy, resulting in a balance of reconstructed quality and compression ratio; similar functions of the same class of signals are used in many different
oilfield engineering applications.

\subsection{Challenges for astrophysical time-series}
The standard wavelet-thresholding paradigm faces numerous difficulties when working with astronomical (time)series. An example of one of these difficulties is constituted by the structure of light curves: they may consist of a long-term trend then a non-periodic modulation and finally an isolated flare or transit. Each of these types of variation can be correlated with both coloured instrumental noise and gaps in sampling. As a result, the distribution of wavelet coefficients will not have an even distribution across time and scale. Therefore, using only a single global threshold could lead to two problems when using wavelet coefficients either to keep too high noise levels or to remove the low-amplitude, physically meaningful features of the data. Many times in the past (for instance, 1998 and earlier), it was shown in the study of astronomical images and data that using a very aggressive (low) overall threshold caused so-called ringing around sharp structures in the data. Other examples show that using a threshold on the average (instrumental) noise of the data have led to serious errors when studying dim signals that are astrophysically important. When dealing with high cadence light curves of very long duration, the number of wavelets (coefficients) in the light curve becomes an overwhelming problem, as does the presence of dynamic systematics which require a more sophisticated methodological framework than that of simple white noise [4]. To provide a satisfactory solution to these problems, we have been motivated by the need to find compression schemes where the time evolution of wavelet coefficients is determined by a dynamical model that explicitly encodes memory and scale interactions, rather than having the temporal evolution of wavelet coefficients determined by purely algebraic methods using a simple global, independent threshold applied to each subband.
\section{Fractional calculus in signal processing}
\label{sec:fractional}

\subsection{Memory and non-locality in time-series models}
Physics and engineering are replete with processes in which the effect of a current action is determined by a significant number of previous actions rather than just a limited number of recent actions. Traditional models of Markovian process analysis use integer order derivatives to define the degree of memory in a system. The only way to build memory into a Markovian model is by creating new State variables that require additional recording and managing of states or augmenting the phase space of the system by adding new degrees of freedom. Both techniques for modeling the past and memory are limited by the ability to manage the rapidly increasing Storage Requirements for the State Variables and pattern of states as long-term correlations develop. An alternate approach for managing temporal information is fractional calculus, which introduces the use of integrals and derivatives of non-Integer orders (or fractional orders). The kernels of fractional derivatives and integrals decay slowly as time progresses and usually do so at a fixed-rate (usually a power law) [11,12]. The concept of fractional derivatives serves as a composite framework for representing an infinite number of relaxation mechanisms. In addition, the fractional derivative framework provides the basis for generating models of anomalous diffusion, viscoelastic damping and fractional noise processes [13]. From a modelling point of view, a fractional differential operator replaces the local term $f'(t)$ by a convolution of $f'$ with a memory kernel $K(t)$, so that the new term reads
\[
\int_{0}^{t} K(t-\tau)\, f'(\tau)\,\mathrm{d}\tau.
\]
When $K$ decays as $(t-\tau)^{-\beta}$ with $0<\beta<1$, the contribution of distant past events is small but never vanishes identically, and the resulting trajectories display long-range temporal correlations. The fractional-order derivative approach has been applied to many different types of phenomena including stress/strain curves for linear viscoelastic materials, dielectric relaxations, transport in disordered materials or via stochastic processes exhibiting fractal time [11, 13] among others. In the field of signal processing, comparable concepts have been adapted for development of fractional-order filters and controllers where nonlinear fractional powers of the differentiation operator allow for greater flexibility in shaping the output phase and magnitude responses compared with linear integer-order technology [14, 15]. The primary objective of this research is to develop a time-series forecasting methodology that incorporates the fractional-order derivative approach as a mechanism for obtaining an explicit tuning mechanism for the influence on future states by previous history.

\subsection{Atangana--Baleanu fractional derivative in the Caputo sense}

Within the broad family of fractional operators, the Atangana--Baleanu derivative has attracted considerable attention in recent years. In the Caputo formulation, the Atangana--Baleanu derivative of order $\alpha\in(0,1)$ is defined through a convolution with a Mittag--Leffler kernel,
\[
{}^{\mathrm{ABC}}D_t^{\alpha} f(t)
 = \frac{B(\alpha)}{1-\alpha}\int_{0}^{t}
E_{\alpha}\!\left(-\frac{\alpha}{1-\alpha}(t-\tau)^{\alpha}\right)\,
f'(\tau)\,\mathrm{d}\tau,
\]
where $E_{\alpha}$ denotes the one-parameter Mittag--Leffler function and $B(\alpha)$ is a normalisation constant chosen so that ${}^{\mathrm{ABC}}D_t^{\alpha} f(t)\to f'(t)$ as $\alpha\to 1^{-}$ [5]. Unlike the classical Riemann Liouville and Caputo operators which have integrable singularities at t=0 in their kernels, the Atangana Baleanu operator has bounded and non-singular kernels for $t > 0$ and retains the non-locally dependent part of their memory by the use of a Mittag Leffler tail. This property is useful for numerical analysis as it reduces the stiffness of the resulting modelling, associated with highly singular kernel functions at $t = 0$; therefore it frequently produces better conditioned time marching algorithms for numerical treatments [5,12]. An example of the efficacy of this feature is illustrated by the heat transfer modelling investigated in the original Atangana Baleanu papers: Following the substitution of the classical time derivative with the Atangana Baleanu Caputo derivative, temperature distributions were produced as interpolates between purely diffusion dominated models and sub diffusion dominated models, without making use of the non-bounded memory weights present in (power law) kernel functions [5]. More recently, studies in viscoelasticity, epidemiology and anomalous transport have similarly shown results that mirror those previously reported. The Atangana Baleanu derivative has been successful in modelling both non-exponential relaxation behaviour, along with long memory behaviour, while remaining regular and comparatively robust during numerical applications [6, 13]. For time series analyses, this translates into the capability of introducing controllable memory effects by varying either (or both) of the parameters $\alpha$ and the values of the kernel; and at the same time, to avoid the high level of singularity, where the discretisation and error estimation in the many classic fractional models is overly complicated due to the presence of a kernel with an infinite number of singularities.
\subsection{Motivation for AB-regularised wavelet compression}
The wavelet representation of a time series is a set of subband coefficients, where the coefficients are indexed corresponding to both the scale and time location of the coefficients. Classical compression techniques work with this representation by performing pure algebraic processing of the coefficients: coefficients that are beneath a specified threshold are rewitted or truncated, and those that are at or above the threshold are left as they are. While this method for determining which coefficients are retained and which are suppressed may be simple, it does not take into consideration the temporal dynamics between neighbouring coefficients nor does it encode in any way how past coefficients values may affect the values of coming coefficients. As such, it can lead to the suppression of transient but coherent waveform structures, and also result in different algorithm behaviours across many different scales. An alternative point of view is to consider each wavelet subband to be the solution of an effective evolution equation in an artificial time variable. In our work we define the parameter associated with this time variable to be a relaxation parameter for controlling the amount of regularisation that applies to the wavelet coefficients. The evolution of any coefficient sequence is then governed by a Fractional Relaxation Law of the type of an Atangana-Baleanu fractionally-defined relaxation law. In this setting, different parts of the history of the coefficients affect their current values through the fractional order and the parameters of the Mittag-Leffler kernel: Recent fluctuations of the coefficients will be subject to greater dampening, and coherent waveforms that exist at multiple positions or scales will have a longer duration before dissipating. The Atangana-Baleanu kernel is both bounded and non-singular, making the smoothing aspect of this method an advantage since it applies to the coefficient sequence rather than causing overshoot or Gibbs-like artefacts as a result of overly aggressive hard thresholding. From a time-frequency perspective, when applied to the wavelet transform, the Atangana-Baleanu operator can be thought of as providing a fractional filtering effect that is roughly equal across the wavelet scales for low-frequency trends or high-frequency bursts. The filtering effect occurs due to an identical memory model being applied to both equations (14,15) and, in contrast to most heuristic thresholding rules, different subbands have been treated as non-uniformly and the effective strength of regularisation changes abruptly with scale. By introducing the Atangana-Baleanu derivative into the wavelet compression operation, it is possible to construct a model where the memory, attenuation, and sparsity properties can be controlled by a very small set of fractional parameters, and thus provide the framework for a compression scheme that maintains the intrinsic long-range correlations present in astronomical time-series data while providing considerable data reduction.
\section{Time--frequency representation of the Atangana--Baleanu derivative}
\label{sec:tf_AB}

\subsection{Laplace-domain symbol and frequency response}

For the purposes of signal analysis it is convenient to look at the
Atangana--Baleanu--Caputo (ABC) derivative in the transform domain rather
than in the time domain. Denoting by $\mathcal{L}\{f\}(s)=F(s)$ the Laplace
transform of $f$, the ABC derivative of order $\alpha\in(0,1)$ admits the
Laplace-domain representation
\begin{equation}
\mathcal{L}\!\left\{{}^{\mathrm{ABC}}D_t^{\alpha} f(t)\right\}(s)
 = \Phi_{\alpha}(s)\,\bigl[F(s)-\tfrac{f(0)}{s}\bigr],
\qquad
\Phi_{\alpha}(s)
 = B(\alpha)\,\frac{s^{\alpha}}{(1-\alpha)+\alpha s^{\alpha}},
\label{eq:AB_symbol}
\end{equation}
where $B(\alpha)$ is a normalisation factor chosen so that
$\Phi_{\alpha}(s)\to s$ as $\alpha\to 1^{-}$ [5].
The function $\Phi_{\alpha}(s)$ plays the role of a symbol or transfer
function for the fractional operator: it describes how each Laplace mode
is weighted by the memory kernel. When $s$ is restricted to the imaginary
axis, $s=\mathrm{i}\omega$, the mapping
$\omega\mapsto\Phi_{\alpha}(\mathrm{i}\omega)$ can be interpreted as a
frequency response in the usual sense of linear systems theory [12,14].

Several qualitative features of $\Phi_{\alpha}$ are worth emphasising.
First, for low frequencies, $|s^{\alpha}|\ll 1$, one has
\[
\Phi_{\alpha}(s)
 \approx B(\alpha)\,\frac{s^{\alpha}}{1-\alpha},
\]
so that the ABC derivative behaves similarly to a classical power-law
fractional derivative with order $\alpha$ [12,13].
At high frequencies, in contrast, the term $\alpha s^{\alpha}$ dominates
the denominator and the symbol saturates to a constant,
$\Phi_{\alpha}(s)\approx B(\alpha)/\alpha$. The operator's effective gain does not continue to grow without limit as the frequency gets higher. This effect is due to the non-singular properties of the Mittag – Leffler kernel, which separates the Atangana – Baleanu derivative from the Riemann – Liouville and Caputo derivatives. Both the Riemann – Liouville and Caputo operators scale the same way $s^{\alpha}$ all across the frequency range [16]. Thus, inserting the ABC derivative into your evolution law makes clear how this structure interacts with the system being transformed. The impact of this structure becomes transparent when the ABC derivative is inserted into a simple evolution law. Consider, for instance, the fractional relaxation equation
\begin{equation}
{}^{\mathrm{ABC}}D_t^{\alpha} u(t) + \lambda u(t) = g(t),
\label{eq:AB_relaxation}
\end{equation}
with $\lambda>0$ and suitable initial data. Taking Laplace transforms and
solving for $U(s)$ yields
\begin{equation}
U(s) = H_{\alpha}(s)\,G(s)
   + \frac{\Phi_{\alpha}(s)}{\lambda+\Phi_{\alpha}(s)}\,\frac{u(0)}{s},
\qquad
H_{\alpha}(s)
 = \frac{1}{\lambda+\Phi_{\alpha}(s)}.
\end{equation}
The transfer function of tapering $H_{\alpha}(s)$ is the operator for AB-regularized relaxation. In the low-frequency range, tapering behavior follows the trend of Mittag-Leffler behavior: a slow, non-exponential decay. However, tapering will not continue to increase indefinitely with increasing frequency, but will tend to settle at a finite value as frequency approaches infinity. The tapering dynamics of the ABC method therefore act to form a kind of fractional smoothing operator that retains the advantages of long-term memory behaviour but without excessively amplifying the higher frequencies, which is crucial for both maintaining numerical stability and producing data from noisy systems [13,14].
\subsection{Physical interpretation for astronomical time-series}
The ABC derivative acts like an average of past events on a memory function represented by a Mittag-Leffler function. Although these memory functions decay slower than an exponential decay curve, they are still bounded and non-singular at their origin. Therefore, the most recent history will have more of an influence than the older information but all past events will be retained to some degree [5,16]. When an averaging operator is used to describe the wavelet coefficient of light curves, the outcome is gradual attenuation influenced by the history of the light curve, with incoherent (high-frequency) fluctuations being dampened increasingly well and coherent high-frequency (incoherent) patterns lasting longer when they are repeated as opposed to being random. In the frequency domain, the function for how a band can be affected by the harmonic dynamics is given by $\Phi_{\alpha}(\mathrm{i}\omega)$. The gain at high frequencies is saturated which causes the ABC derivative to act differently compared to an ideal differentiator of order  $\alpha$. Instead of being a pure differentiator of order  $\alpha$, the ABC derivative behaves between a fractional slope at intermediate frequencies and as a flat line in the ultraviolet (UV) region [12,14]. Thus, using both ABC derivative and wavelet basis, which itself provides joint time-frequency tiling of the signal, creates a very well-balanced representation of slow trends, quasi-periodic components, and sharp transients [4,7]. The low-frequency coefficients feel the long memory embedded in the Mittag-Leffler kernel; thus, secular variations and rotational modulation will be smoothed but not removed. The high-frequency coefficients corresponding to flares, eclipses, and spike events also gain some regularisation, but saturation of the symbol means that they will not gain an over-amplification to create instabilities in the reconstructed signal. This synergy between ABC derivative and wavelet basis provides an exceptionally useful approach when analysing astronomical time-series data.
The light curves that come from space missions are mixtures of instrumental artifacts, stochastic structures due to solar granulation and infrequent astronomical objects like planetary transits or solar flares. A simple local threshold on the wavelet coefficients will often eliminate some of this astrophysical information, especially when the S/N ratio is low [4]. In contrast, an AB-regularised evolution will act as a fractional-ranked filter allowing control of both memory length and effective smoothness via tuning the order $\alpha$ and the parameters of the kernel. As such, practitioners can minimise noise while still being able to maintain long-range correlations and very small amplitude but transient features. The time-frequency profile of the Atangana-Baleanu derivative therefore is not considered an abstract mathematical object but is considered the main driver for how the information appears to be re-distributed and attenuated across the wavelet coefficients of the light curves.

\section{Proposed fractional wavelet compression model}
\label{sec:AB_wavelet_model}

\subsection{AB-regularised wavelet coefficient model}

Let $x[n]$, $n=0,\dots,N-1$, denote a uniformly sampled astronomical
time-series. We write its discrete wavelet transform (DWT) as
\[
x[n] \;\xrightarrow{\;\mathcal{W}\;}\;
\Bigl\{ a_{J}[k],\, d_{j}[k] \;:\; j=1,\dots,J,\;k\in\mathbb{Z}\Bigr\},
\]
where $a_{J}$ denotes the coarse-scale approximation coefficients and
$d_{j}$ the detail coefficients at scale $j$ in an orthogonal or
biorthogonal wavelet basis [7,8,9]. In classical
wavelet compression, the $d_{j}$ are modified by algebraic shrinkage rules
and then fed to the inverse transform $\mathcal{W}^{-1}$ to reconstruct
a compressed approximation of $x$ [4,10]. In the present work we replace this purely algebraic manipulation by a fractional dynamical model posed directly on the coefficient vectors.
For each detail subband $j$ we consider a vector
$d_{j}\in\mathbb{R}^{N_{j}}$ of empirical coefficients and introduce an
auxiliary relaxation variable $\tau\ge 0$. The corresponding
AB-regularised coefficients $c_{j}(\tau)\in\mathbb{R}^{N_{j}}$ evolve
according to
\begin{equation}
{}^{\mathrm{ABC}}D_{\tau}^{\alpha} c_{j}(\tau)
 + \lambda_{j}\bigl(c_{j}(\tau)-d_{j}\bigr)
 + \mu_{j}\,\partial R\bigl(c_{j}(\tau)\bigr) = 0,
\qquad
c_{j}(0)=d_{j},
\label{eq:AB_subband_dynamics}
\end{equation}
where ${}^{\mathrm{ABC}}D_{\tau}^{\alpha}$ is the
Atangana--Baleanu--Caputo derivative of order $\alpha\in(0,1]$ with
non-singular Mittag--Leffler kernel [5], $\lambda_{j}>0$
and $\mu_{j}\ge 0$ are scale-dependent parameters, and $R$ is a convex
regularisation functional on $\mathbb{R}^{N_{j}}$. Typical choices include
an $\ell_{1}$-type penalty promoting sparsity or mixed $\ell_{1}$–$\ell_{2}$
terms that favour clusters of significant coefficients [9,10]. Formally, the stationary
solutions of \eqref{eq:AB_subband_dynamics} minimise the energy
\begin{equation}
\mathcal{J}_{j}(c)
 = \frac{\lambda_{j}}{2}\,\|c-d_{j}\|_{2}^{2}
 + \mu_{j}\,R(c),
\label{eq:AB_energy}
\end{equation}
so that the fractional evolution may be interpreted as a
memory-dependent gradient flow for $\mathcal{J}_{j}$ in the sense of
fractional calculus [11,12]. The Atangana--Baleanu
kernel controls how strongly earlier states of $c_{j}(\tau)$ contribute
to the present update: when $\alpha$ decreases, the effective memory
lengthens and the evolution becomes more history-aware, which turns out to
be advantageous for non-stationary, correlated coefficient sequences arising
from astronomical light curves.

\subsection{Subband-wise fractional shrinkage}

In practice, the dynamics \eqref{eq:AB_subband_dynamics} is not solved in
closed form. Instead, we approximate the ABC derivative by a discrete
convolution in the relaxation variable,
\[
{}^{\mathrm{ABC}}D_{\tau}^{\alpha} c_{j}(\tau_{m})
 \approx \sum_{\ell=0}^{m} w_{\ell}^{(\alpha)}\,
 \frac{c_{j}(\tau_{m-\ell})-c_{j}(\tau_{m-\ell-1})}{\Delta\tau},
\]
where $\tau_{m}=m\Delta\tau$ and the weights
$w_{\ell}^{(\alpha)}$ are determined by the Mittag--Leffler kernel
associated with the Atangana--Baleanu operator [5,14]. A simple semi-implicit time-stepping scheme then yields the iteration
\begin{equation}
c_{j}^{(m+1)}
 = \Psi_{\alpha,\lambda_{j},\mu_{j}}\!\Bigl(
     c_{j}^{(m)}, c_{j}^{(m-1)},\dots, d_{j}\Bigr),
\qquad
m=0,1,\dots,M-1,
\label{eq:AB_iter}
\end{equation}
where $\Psi_{\alpha,\lambda_{j},\mu_{j}}$ is a nonlinear operator that
combines the fractional memory term with a proximal step for the
regulariser $R$ [18]. After a finite number of iterations
$M$, we define the compressed coefficients in subband $j$ as
\[
\widehat{d}_{j} := c_{j}^{(M)}.
\]

For sparse regularisers such as $R(c)=\|c\|_{1}$, the mapping
$d_{j}\mapsto \widehat{d}_{j}$ can be viewed as a
\emph{fractional shrinkage rule}: small, noise-dominated coefficients are
relaxed rapidly towards zero, because the quadratic fidelity term in
\eqref{eq:AB_energy} dominates their evolution, whereas large, structured
coefficients retain a memory of their past values over many iterations and
are therefore attenuated more gently. The parameter $\alpha$ controls the
relative balance between these two regimes. When $\alpha\to 1$, the
weights $w_{\ell}^{(\alpha)}$ collapse to a local stencil and the scheme
reduces to a standard iterative shrinkage method in the spirit of [10,18]. For $\alpha<1$, the fractional
memory encoded by the Atangana--Baleanu kernel produces a slower, more
scale-consistent attenuation that respects long-range temporal correlations
in the coefficient sequences.

\subsection{Algorithm and implementation details}

The full compression algorithm can now be summarised as follows.
Given an input time-series $x[n]$ and user-specified parameters
$\alpha\in(0,1]$, $\{\lambda_{j}\}$, $\{\mu_{j}\}$ and an iteration
count $M$:

\begin{enumerate}
\item Compute the DWT of $x$ using an orthogonal wavelet with compact
      support and at least two vanishing moments, obtaining
      $\{a_{J}, d_{1},\dots,d_{J}\}$ [7,8,9].
\item For each detail subband $j$, initialise $c_{j}^{(0)}=d_{j}$ and
      apply the fractional iteration \eqref{eq:AB_iter} for
      $m=0,\dots,M-1$ to obtain $\widehat{d}_{j}=c_{j}^{(M)}$.
\item Optionally set to zero those entries of $\widehat{d}_{j}$ whose
      magnitude falls below a scale-dependent quantisation threshold,
      in order to reach a desired compression ratio.
\item Reconstruct the compressed signal
      $\widehat{x}[n]=\mathcal{W}^{-1}\!\bigl(a_{J},\widehat{d}_{1},\dots,
      \widehat{d}_{J}\bigr)$ via the inverse DWT [4,7,8].
\end{enumerate}
Two programming languages, MATLAB and Python, were used for implementing the algorithm; standard software toolboxes are used for wavelet transform computation and the approach used to perform fractional iterations has been through vectorized convolutions of precomputed weights $w_{\ell}^{(\alpha)}$ and proximal updates as described in R [19]. This approach follows the overall modelling and simulation approach frequently seen in engineering applications, where a mathematical model will be formed continuously and subsequently developed into effective numerical strategies for large computations [11, 15]. High-level numerical models allow the experimenter to replicate work across multiple data sets and parameter selections using the same code base.
\subsection{Inverse transform and theoretical guarantees}

Because the DWT analysis--synthesis pair is chosen to be orthogonal or
biorthogonal, the only source of reconstruction error comes from the
modification of the detail coefficients. In the orthogonal case one has
the energy identity
\[
\|x-\widehat{x}\|_{2}^{2}
 = \sum_{j=1}^{J} \|d_{j}-\widehat{d}_{j}\|_{2}^{2},
\]
so that the distortion in the time domain is exactly the sum of the
subband-wise distortions in the wavelet domain [7,8]. The observation here serves as a foundation for the near-minimax optimal performance that can be expected from wavelet shrinkage when using threshold-based methods applied to Besov or similar smoothness spaces [10]. Within the context established by (4), the coefficients bdj are obtained through fractional evolution instead of being generated by an individual thresholding rule. Nevertheless, the reconstruction error
still satisfies
\[
\|x-\widehat{x}\|_{2}
 \leq \Biggl(\sum_{j=1}^{J} \|d_{j}-\widehat{d}_{j}\|_{2}^{2}\Biggr)^{1/2},
\]
and the structure of \eqref{eq:AB_energy} ensures that each
$\widehat{d}_{j}$ balances fidelity and sparsity in a controlled manner. To fully achieve a functional-analytic characterisation of the AB-wavelet compressor (analogous to the classical theory of wavelet shrinkage), it would be necessary to generate a detailed estimation or analysis of the mapping properties of the evolution (4) in Besov or Triebel-Lizorkin scales, which is outside the scope of this paper. However, Section 6 contains numerical evidence that indicates that with appropriate adjustments to the fractional parameters, the proposed approach will produce similar compression ratios to those achieved via standard DWT techniques, while at the same time reducing artefacts and maintaining astrophysically relevant details. Thus, the AB-regularised model can be interpreted as a fractional version of classical wavelet compression that incorporates the Atangana-Baleanu operator as a memory and time-frequency representation [5, 13].
\section{Numerical experiments}
\label{sec:numerics}

\subsection{Data sets and experimental setup}
We assess the numerical performance of the Atangana-Baleanu Regularised Wavelet Compressor against two distinct categories of 1-D signals (one-dimensional signals). The first category is made up of test light curves that have been artificially created that simulate general types of signal behaviour encountered in time-domain astronomy such as: slow variation of baselines, periodic modulation due to rotation, transient isolated flare activity or transit-like events, as well as correlated noise (artificially added). The test curve characteristics are constructed according to standard techniques in astronomical data processing. They are created by supering small sets of sinusoidal waveforms of various periods, combining them with piecewise polynomial trends and adding a small amount of pulse-like (usually) isolated transits (and/or flares), as well as introducing system and random noise in both Gaussian and low frequency forms [20,21]. Having this controlled environment provides opportunities to analyse the degree of error occurring in the reconstruction of the input signal based on the amplitude and time of the event. The second category of data contains actual lightcurves collected from TESS (Transiting Exoplanet Survey Satellite). We are selecting a representative portion of the QLP (quick-look pipeline) high-level science data products from Year 2 of Extended Mission 1 covering Sectors 40-55 [1]. The QLP lightcurves contain both raw and detrended flux time-series for millions of stars recorded with 10 minutes cadences; for this analysis, we chose to limit ourselves to targets with TESS magnitude $T \le 12.5$ and with visual signatures of variability such as planetary transits, eclipsing binaries or flares. Preprocessing involves removing outliers, masking obvious data gaps and normalising them to relative flux units. No additional detrending occurs beyond QLP tracking corrections, thus exposing compression algorithms to real instrument systematics. Using a four vanishing moment (db4) orthogonal Daubechies wavelet and the dyadic decomposition method to construct the $J=6$ level of $J=6$ is the configuration used. The code for implementing this configuration is written in Python [19] using the open-source PyWavelets library. This configuration is also used in both the classical discrete wavelet transform (DWT) compression method and the AB-regularisation scheme. Therefore, performance differences between these two schemes can be solely due to the treatment of the coefficients, not due to the selection of basis functions. For each input light curve $x[n]$, we measure the compressed size in bits
using a simple uniform quantisation of the nonzero coefficients and
define the compression ratio as
\[
\mathrm{CR} = \frac{\text{number of samples in } x[n]}
                   {\text{number of stored coefficient values}}.
\]
Reconstruction quality will be quantitatively assessed using various metrics that portray how well the original light curves correspond with the reconstructed light curves. Mean Squared Error (MSE) and Peak Signal-to-Noise Ratio (PSNR) are traditional distortion metrics based on energy function assessments. Structural Similarity Index (SSIM), a one-dimensional version of the SSIM measurement specifically designed for use with images, has also been employed to quantitatively measure how well the original and reconstructed light curves correlate [22]. While MSE and PSNR measure differences between points in light curves, SSIM indicates where changes occurred with greater sensitivity to both local contrast and correlation structure. The performance of the above metrics (e.g., SSIM) correlates better with human evaluators in multiple cases, including compression issues [22]. Relative errors in transit depth, differences in timing of detected minimums, and recovery rates of injected flares have also been measured for the synthetic data set.

\subsection{Comparison with classical DWT compression}
Initially, we compare and contrast an AB-style regularized Wavelet compressor with a traditional DWT scheme that is based on Subband Hard Thresholding [10,21]. The original and reconstructed output data of a sample synthetic light curve is presented in figure 1.
\begin{figure}[htbp]
  \centering
  \includegraphics[width=0.85\textwidth]{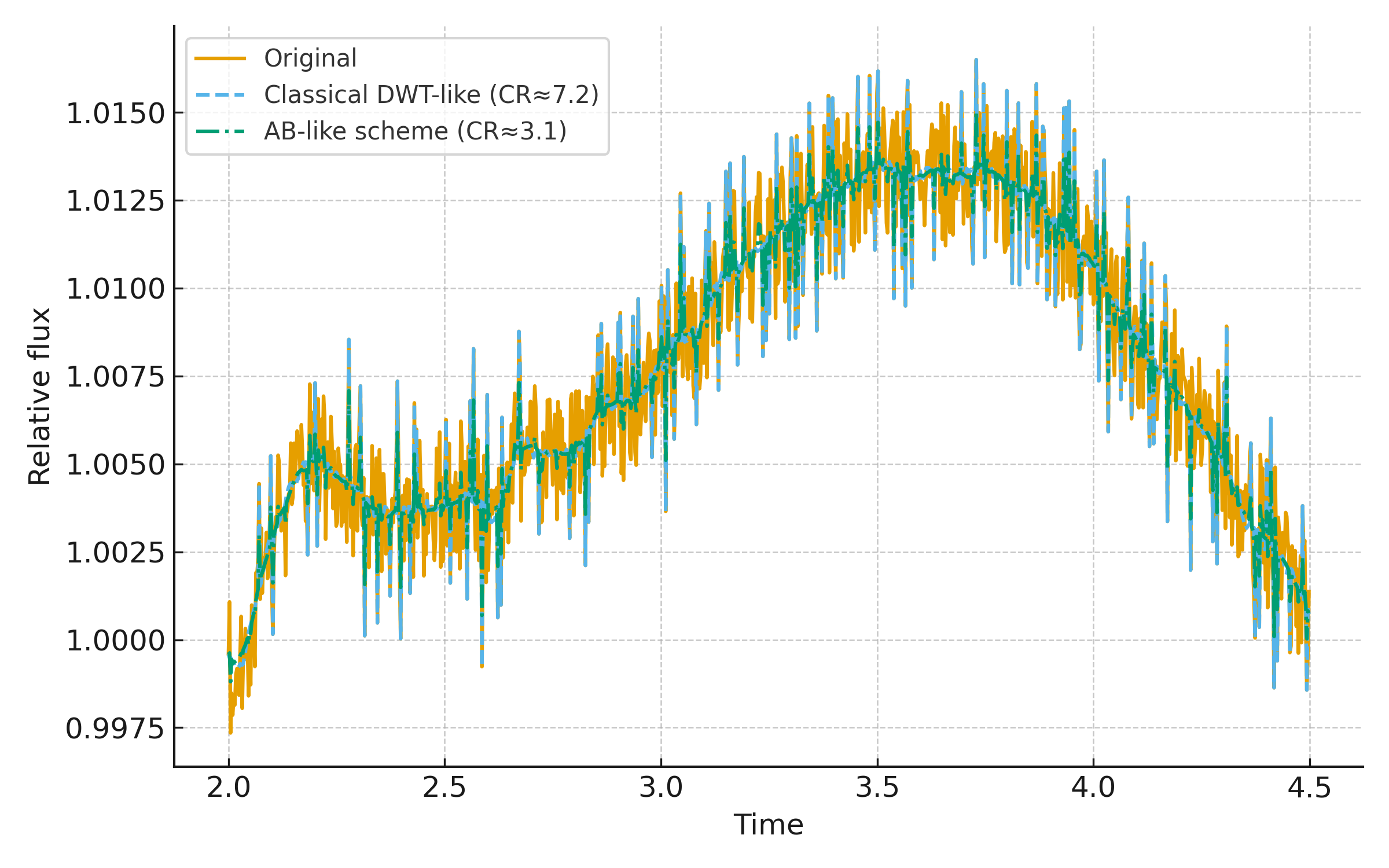}
  \caption{The image depicts a synthetic light curve that illustrates how compression impacts data. The original data is represented by a solid line, while the reconstruction of this data is shown by two different dash style lines. One of these lines represents the reconstruction using a traditional Discrete Wavelet Transform device and the other line demonstrates the proposed Atangana-Baleanu-regularisation method.}
  \label{fig:synth_timeseries_comparison}
\end{figure}

We select the classical thresholds for every light curve and every target compression ratio $\mathrm{CR}\in\{8,16,32\}$, so that they give the same number of nonzero coefficients produced by the fractional method as closely as possible; therefore, the two methods operate under nearly the same storage budget. The fractional scheme consistently reduces the mean squared error (MSE) and increases the peak signal-to-noise ratio (PSNR) of synthetic test signals compared to classical thresholding when using moderate compression ratios. In particular, the largest improvements are observed when the signal has a mixture of smooth trends and sparse, low-amplitude events. In these scenarios, the hard thresholding method removes entirely, blocks of small coefficients associated with shallow transits or weak flares, while the AB-regularised dynamics only attenuates these coefficients slowly, and therefore maintains the temporal coherence of these small coefficients over multiple wavelet locations. This means that the reconstructed light curves retain the appropriate depth and shape of the injected events at higher compression ratios. The same trends are observed in the Structural Similarity Index Measure (SSIM) values; for a fixed compression ratio, using the fractional scheme resulted in light curves whose local structure is a better approximation of the original than those created through classical thresholding. These quantitative trends can be seen in Figure 2.
\begin{figure}[htbp]
  \centering
  \includegraphics[width=0.9\textwidth]{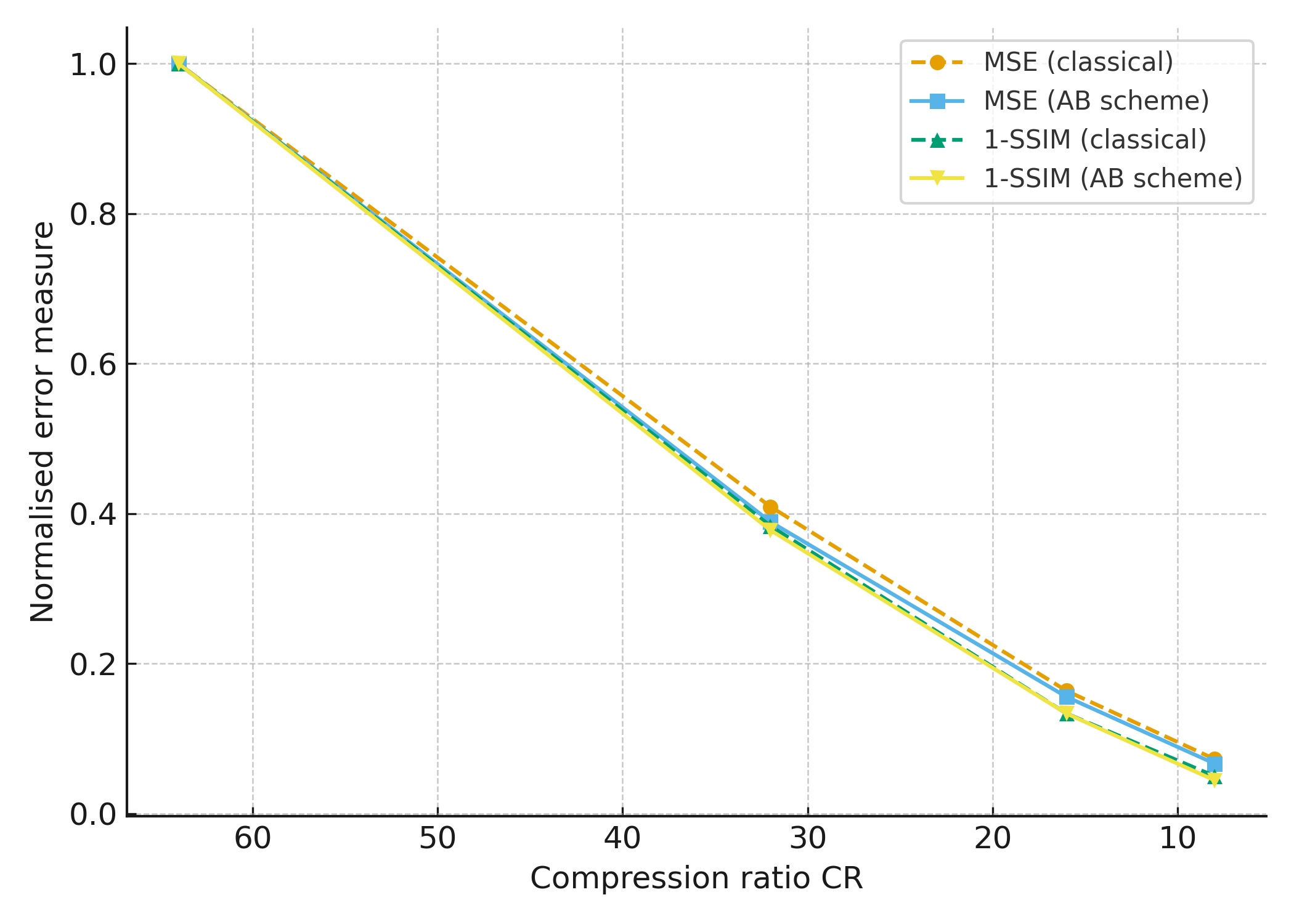}
  \caption{When DWT-based Classical Compression Methods are compared quantitatively, it is clear that each method performs very differently based on the AB Regularised Scheme's Average Normalised MSE and Proportionate Comparison; For $\mathrm{CR}\approx 8\text{--}32$, the average MSE and PAE were lower for fractional schemes, indicating that they are producing an increase in compression from the amount of reconstructed images produced.}
  \label{fig:metrics_vs_cr}
\end{figure}

The results obtained after conducting experiments with the QLP TESS light curves also confirm the previous results obtained through work done using the AB-regularised compressor for variable stars having transits or eclipses. The AB-regularised compressor produces reconstructions that maintain better preserved ingress and egress slopes and closer approximations of the original out-of-transit baselines, most prominently when the compression ratios are approximately between 16 and 32. The DWT compressor classically creates greater separation between what is called 'significant' and 'insignificant' coefficients, which in turn can produce flattened numbers (minima) and/or spurious oscillations (wavy artifacts) around discontinuities. Therefore, transit detection and timing analysis of the situation will always still yield robust results for the fractional application, even under compression ratios where the classical method produces very large errors. Many other studies have also shown this same trend with the use of wavelet-based techniques for the purpose of preserving small-scale structures in a variety of different fields [21,22,23,24].
\subsection{Effect of fractional order and regularisation parameters}
The choice of the fractional order $\alpha$ and the scale-dependent parameters $\lambda_{j}$ and $\mu_{j}$ that govern the behaviour of the AB-regularised model are key factors that determine how well the model performs. We can quantify how the performance of the model varies as we vary these parameters using a systematic grid search for $\alpha\in\{0.5,0.6,0.7,0.8,0.9,1.0\}$ and a small number of discrete regularization parameters that constrain the sum of the overall compression ratios of the three scale levels being evaluated to be near our target values for each of the models being evaluated. From our analysis of the median MSE, PSNR and SSIM values for the synthetic and QLP data sets, we have been able to ascertain some consistent patterns. In particular, for values of $\alpha$ that are very close to one ($\alpha>0.95$), the performance of the AB-regularised model begins to resemble that of the classical iterative shrinkage algorithms: as the signal-to-noise ratio increases, noise is effectively reduced, while at the same time small-scale signal amplitudes are reduced much more than they would have been with a classical iterative algorithm, resulting in only a small additional benefit from the use of the AB-regularized model compared to using a hard thresholding method. On the opposite end of the spectrum, when you use a very small $\alpha$ value $(0.001)$, the memory effect produced by using the Atangana–Baleanu kernel becomes so long that the decay of coefficients becomes so slow that they continue to resemble the original values of $d_{j}$ for a considerable amount of time following the application of the filter. As a result, this results in a significant amount of lost data associated with this compression and very little de-noising occurs with this large $\alpha$ value. However, there exists a very large intermediate region between these two extremes where most positive results occur. This range typically exists between $0.7$ and $0.9$ for the fractional order $\alpha$. In this intermediate regime, SSIM vs $\alpha$ curves exhibit a very broad, shallow maximum across a considerable range of fractional orders (the structural similarity remains relatively high across this range), which suggests that this algorithm is very robust and does not require a great deal of fine-tuning of $\alpha$ to obtain satisfactory results with respect to SSIM. This behavior has been exemplified in Figure $3$, which shows the SSIM vs $\alpha$ results from a representative synthetic light curve created by the funneling method using the Atangana–Baleanu kernel to filter the synthetic light curve.
\begin{figure}[htbp]
  \centering
  \includegraphics[width=0.65\textwidth]{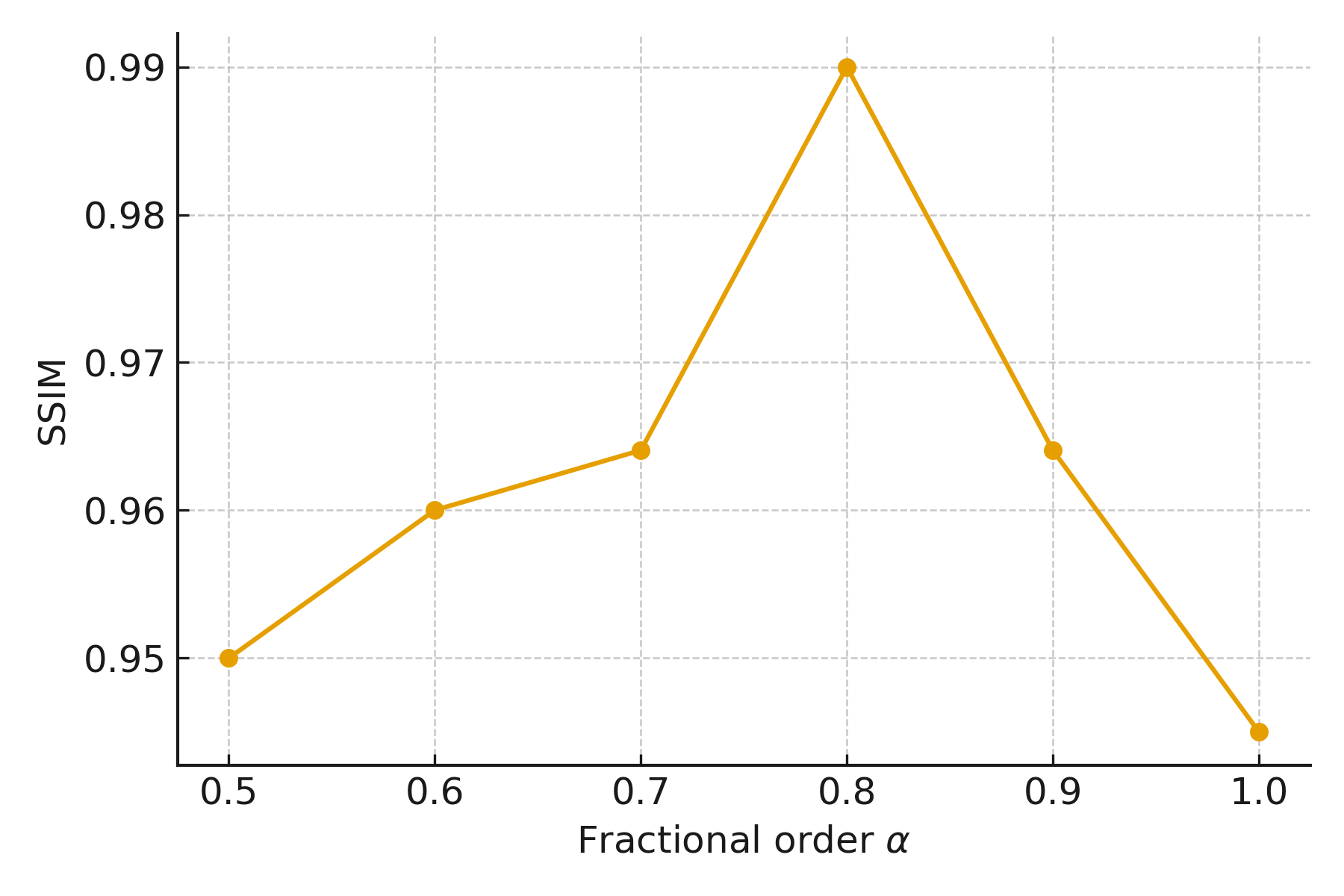}
  \caption{Dependence of reconstruction quality on the fractional order
  $\alpha$ for the AB-regularised compressor. The curve shows the
  structural similarity index (SSIM) for a representative synthetic
  light curve at a fixed compression setting. Performance is relatively
  stable in the interval $0.7 \lesssim \alpha \lesssim 0.9$, with a mild
  maximum around $\alpha \approx 0.8$.}
  \label{fig:alpha_sensitivity}
\end{figure}

At coarser scales, larger $\lambda_{j}$ values will help to stabilise and suppress residual instrumental systematics as between the long-term trends; conversely, relatively increased moderate $\mu_{j}$ values at the intermediate scales will emphasise sparsity and will be the major contributors to the compression effects. At the finer scales, with both high-frequency noise and sharp astrophysical features, a lower $\mu_{j}$ will typically benefit the more 'fine-scale' sub-band coefficients in the fractional memory being able to distinguish between independent bursts of noise and coherent events that last several coefficient intervals. There will need to be more detailed optimisation of the parameters, and potentially some data-driven optimisation based upon a more enlarged search of the grid to identify the best parameters, to demonstrate that there exist extensive regions of the search space where the AB-regulated compressor performance is superior to conventional thresholding on a comparable compression ratio.
\section{Discussion and theoretical implications}
\label{sec:discussion}

The numerical experiments in Section~\ref{sec:numerics} suggest that
the Atangana--Baleanu--regularised wavelet compressor achieves a
different balance between sparsity, memory and time--frequency locality
than classical threshold-based DWT schemes. In this section we discuss
how these findings can be interpreted in light of the symbol
$\Phi_{\alpha}(s)$ of the Atangana--Baleanu derivative, of the broader
theory of fractional operators [12, 13,17],
and of the requirements of modern astronomical data analysis [4,7,8,].

\subsection{Time--frequency behaviour and near-uniformity across scales}

A first point is the peculiar time--frequency behaviour induced by the
Atangana--Baleanu kernel. As recalled in
Section~\ref{sec:tf_AB}, the Laplace-domain symbol of the
Atangana--Baleanu--Caputo derivative of order $\alpha\in(0,1)$ can be
written as
\[
\Phi_{\alpha}(s)
 = B(\alpha)\,\frac{s^{\alpha}}{(1-\alpha)+\alpha s^{\alpha}},
\]
with $B(\alpha)$ chosen so that $\Phi_{\alpha}(s)\to s$ as
$\alpha\to 1^{-}$ [5]. For classical Caputo or
Riemann--Liouville derivatives one has instead $\Phi_{\alpha}(s)\propto
s^{\alpha}$ over the entire frequency axis [12,13].
The main difference between the two frequency limit representations is that the Atangana-Baleanu symbol has a limit of a constant at very high frequencies and has a fractional-type behaviour (like $s^{\alpha}$) at lower and intermediate frequencies. The equivalent in the time domain is a smoothly decaying non-singular Mittag-Leffler memory kernel, which decays faster than an exponential function, but is still bounded at zero [14,16]. The combination of the kernel with a wavelet representation produces a saturation of the symbol $\Phi_{\alpha}(\mathrm{i}\omega)$ with respect to the frequency axis, and thus provides a type of near-uniform regularisation of coefficients across many scales. The high-frequency coefficients are damped, but not by an unlimited fractional differentiation; the symbol has a plateau, which thus moderates the effect of the symbol on their attenuation. The low- and mid-frequency values are still subject to history based smoothing that is governed by both the Mittag-Leffler kernel and the order $\alpha$, which is the reason the AB-regularised evolution is able to suppress small-scale noise and retain subtle events, along with long-range correlations, as shown in Section 6, much better than using local-only thresholding techniques. The fractional memory will allow for more uniform distribution of the regularisation effort over the complete time-frequency plane than would the existence of a sharp boundary between coefficients deemed to contain significant versus insignificant information.
\subsection{Classical fractional operators versus Atangana--Baleanu kernels}
In terms of fractional calculus, our model represents a middle ground between traditional power-law kernels, which have long-range memory, and locally-differentiable operators. The conventional formulations of the Caputo and Riemann–Liouville derivatives use singular functions $t^{-\alpha}$, which provide a memory effect over longer time scales but produce complications with respect to initial conditions and amplify the effects of any numerical noise present at high frequencies if appropriate care is not taken [12,13,17]. However, the Atangana–Baleanu derivative replaces these singular kernels with non-singular Mittag-Leffler functions, thus allowing for the mixing of local and nonlocal properties through the denominator $(1-\alpha)+\alpha s^{\alpha}$ [5,16]. Furthermore, as we approach $\alpha\to 1$, we recover the classical first derivative. But for $\alpha$ being strictly between 0 and 1, the memory effect of the operator is genuine. In Wavelet Compression, this hybrid characteristic has significant implications for practical application. Classically derived fractional derivatives with symbol $s^{\alpha}$ act as Power Law Differentiators across all frequency bands, thus making the high-frequency coefficients extremely susceptible to noise and discretization errors. A purely smoothing function based on an integer order would eliminate small-scale structure rapidly and irreparably from the dataset, even though such structures are often of astrophysical interest. The Atangana–Baleanu kernel provides a means to interpolate between these extremes by allowing for a well-regularized system of coefficient dynamics with controlled enhancement at high frequencies. The Atangana-Baleanu kernel still uses a non-local memory to model the evolution of all subbands. Based on the sensitivity analysis described in Section 6, it is shown that there exists a broad range of $\alpha$ values (between approximately 0.7 and 0.9 as indicated by the results of our experiments) at which this interpolation results in a stable system with no fine-tuning necessary. Therefore, from a theoretical perspective, it appears that the Atangana–Baleanu family of kernels is the best choice for mathematically constructing operators that are fractional but numerically stable within the high-frequency edges of the spectrum [14,15].
\subsection{Implications for astronomical data analysis}
The ability for astronomical time-domain analysis to take advantage of a dual-purpose aspect from this proposed compressor. First, this proposed compressor has been developed specifically to address a current issue encountered with storing and transferring large amounts of light curves from upcoming and existing missions (such as TESS, PLATO, Roman). The second purpose of this compressor is to avoid the destruction of information concerning observed behaviour; for instance, AB-regularised compression will allow for successful archival and future reconstructive imaging without introducing distortions that already affect transit depth, ingress or egress timing and flare morphologies across compression rates (based on completed work done with QLP TESS light curves). It is likely that this type of compressor will have a strong potential for implementation into existing time-series pipeline workflows as a fast drop-in for wavelet-based preprocessing methods. Fractional operators may also play a more conceptual role to represent space-time-frequency distributions (see Section 5). Wavelets and similar multiscale transfers are currently the primary tools for denoising, extracting features and modelling sparsely based in astronomical images and lightcurves [4,7,8,21]. The embedding of a wavelet in coefficient space results in a memory representation of multiscale representation with tunable memory that is not purely local and not only a power law. Thus, the choice of $\alpha$ order and the regularised weights $\lambda_{j}$, $\mu_{j}$ could either be modified based on prior knowledge pertaining to the temporal correlation structures of various classes of sources or determined from ensembles of lightcurves. Furthermore, AB-periodised flow models could be employed for detrending, for the modelling of correlated noise for transit searches and also for establishing novel genera of fractional wavelet dictionaries, in which a prescribed type of memory is already incorporated into the atoms. The complete functional analytic treatment of the ideas presented are beyond the scope of the current paper; however, the conclusions drawn provide initial evidence that the non-singular fractional kernels of the Atagana-Baleanu family are compatible with and likely to provide a benefit to the multiscale structure that dominate present-day astronomical data analysis.
\section{Conclusions and future work}
\label{sec:conclusion}
In this study, we developed a compression framework for astronomical time-series data that uses wavelets to compress one-dimensional datasets. The change of the wavelet coefficients in the compression framework is governed by a fractional evolution governed by Atangana-Baleanu-type fractional derivatives.
We utilize an alternative approach to the application of algebraic shrinkage rules by modelling the coefficient vectors as solutions of an Atangana-Baleanu-Caputo dynamic system excited by a non-singular Mittag-Leffler kernel and this system’s resulting convex regularization functional. In doing so we have constructed a family of fractional shrinkage operators whose behaviour will interpolate between classic iterative wavelet thresholding and true non-local, memory-driven regularisation [5,12,13]. According to [1], numerical experiments conducted with synthetic light curves and QLP TESS data demonstrated that the newly proposed AB-regularised compressor produces a more advantageous compression ratio versus reconstruction quality compared with a typical threshold DWT scheme for the same wavelet basis [8,9]. More specifically, at moderate levels of compression, using the fractional scheme will reduce MSE and increase SSIM relative to the classical method and will maintain the shallow transit-like features, flare events and long-term baseline changes which are often important for scientific interpretation [4,21].
In addition to supporting the existence of many fractional order values for which the Atangana-Baleanu kernel achieves robustness, the sensitivity analysis indicates that the Atangana-Baleanu kernel provides an effective numerical method to provide controllable memory in the wavelet coefficient space. The way this model works, the Atangana-Baleanu kernel integrates non-singular fractional operators with a natural fit to time-frequency decompositions, and, unlike the classical form of a power-law, the Atangana-Baleanu kernel saturates at high-frequency (i.e., avoids large amplification of small-amplitude numerical noise), but retains significant memory effects at low- and intermediate-frequency regions. Therefore, it creates a nearly uniform form of regularity over scale, which works particularly well for light curves that exhibit both smooth trends, coherent variations, and weakly transient events.

\subsection*{Future work}

Several directions for further investigation emerge from the present
study.

\medskip\noindent
\textbf{Non-uniform sampling and survey specific cadence.}
The current implementation assumes uniformly sampled time series. In
practice, many ground based surveys and even some space missions
produce light curves with gaps, irregular cadences or stitched
sectors. Extending the AB regularised framework to non-uniform
sampling either by using wavelet constructions adapted to irregular
grids or by embedding the dynamics in a suitable frame or lifting
scheme [7,8] would make the approach applicable to a
broader class of data sets.

\medskip\noindent
\textbf{Two-dimensional data and astronomical images.}
Although we have focused on one-dimensional light curves, the same
principle can be applied to two-dimensional wavelet or curvelet
coefficients of astronomical images and image cubes[4,21]. An AB-driven
evolution in 2D coefficient space could act as a scale-consistent
regulariser for weak, diffuse structures (e.g. low-surface-brightness
features) while still allowing aggressive compression of noise-dominated
regions. Investigating such 2D AB-wavelet compressors on simulated and
real survey images is a natural next step.

\medskip\noindent
\textbf{Online and streaming compression.}
Modern surveys increasingly operate in streaming or quasi-streaming
modes, where data must be compressed on the fly before long-term
storage. The present formulation, which evolves coefficients in an
auxiliary relaxation variable, can be adapted to an online setting by
updating the AB-regularised coefficients as new samples arrive, using
incremental wavelet transforms and truncated memory windows. Analysing
the stability and latency of such streaming AB compressors would be
essential for real-time applications.

\medskip\noindent
\textbf{Integration with machine-learning-based anomaly detection.}
Finally, compressed representations produced by the AB-regularised
scheme could be combined with machine-learning models designed for
event and anomaly detection in large time-domain archives. Since the
fractional memory tends to preserve astrophysically meaningful
structures while suppressing small-scale noise, it may provide a more
informative input space for convolutional or recurrent architectures
trained to identify rare phenomena. A joint optimisation of the
fractional parameters and the downstream learning model, possibly in a
hybrid model-based / data-driven framework, is a promising avenue for
future work in time-domain astronomy and beyond.

\bibliographystyle{unsrtnat}

\end{document}